\documentclass[12pt]{iopart}
\usepackage{iopams}  
\usepackage{graphics}
\usepackage{epsfig}
\eqnobysec
\def\lambdabar {\mathchar'26\mkern-10mu\lambda}

\begin{document}

\title[A relativistic model of the $N$-dimensional singular oscillator]{A relativistic model of the $N$-dimensional singular oscillator}

\author{S.M. Nagiyev\dag, E.I. Jafarov\dag  
\footnote[3]{To whom correspondence should be addressed
(azhep@physics.ab.az)} and M.Y. Efendiyev\ddag}

\address{\dag\ Institute of Physics, Azerbaijan National Academy of
Sciences, Javid av. 33, AZ1143, Baku, Azerbaijan}

\address{\ddag\ Azerbaijan Cooperation University, Narimanov av. 86, AZ1106, Baku, Azerbaijan}

\begin{abstract}

Exactly solvable $N$-dimensional model of the quantum isotropic singular oscillator in the relativistic configurational $\vec r_N$-space is proposed. It is shown that through the simple substitutions the finite-difference equation for the $N$-dimensional singular oscillator can be reduced to the similar finite-difference equation for the relativistic isotropic three-dimensional singular oscillator. We have found the radial wavefunctions and energy spectrum of the problem and constructed a dynamical symmetry algebra.
\end{abstract}


\submitto {\JPA}


\section{Introduction}

\label{int}

The search for exactly-solvable quantum-mechanical systems is one of the interesting concepts in theoretical and mathematical physics. It is well known that the Schr\"odinger equation can be exactly solved only for specific potentials. Its exact solution history can be traced back to the days of exact solution of the non-relativistic linear harmonic oscillator and Coulomb interaction problems \cite{landau,moshinsky}.

The non-relativistic quantum singular oscillator is also one of the exactly solvable problems studied in detail due to its considerable physical interest and a lot of applications \cite{camiz,chumakov,hartmann,nieto}. In the non-relativistic quantum mechanics the singular oscillator is trivially related to the linear harmonic oscillator. However, this is no longer true in the relativistic theory.

Recently we constructed a relativistic model of the quantum linear singular oscillator and showed that the approach used there can be generalized for construction of the three-dimensional isotropic singular oscillator model. The purpose of the present paper is to apply the results obtained in \cite{nagiyev1,nagiyev2} to the $N$-dimensional case.

In section 2, we briefly discuss the finite-difference relativistic quantum mechanics. Its generalization to the $N$-dimensional case is considered in section 3. We propose a relativistic finite-difference model of the isotropic $N$-dimensional singular oscillator and find its wavefunctions and energy spectrum in section 4. In section 5, we constructed a dynamical symmetry algebra of the model under consideration. Conclusions are given in section 6.

\section{The finite-difference relativistic quantum mechanics}

Our $N$-dimensional model is formulated in the framework of the finite-difference relativistic quantum mechanics \cite{kadyshevsky,kadyshevsky2,freeman,kadyshevsky3,kadyshevsky4,jhung,milton,frick}. The finite-difference relativistic quantum mechanics possesses many important features of the non-relativistic quantum mechanics, but unlike non-relativistic quantum mechanics here the wavefunction of the relative motion satisfies a finite-difference Schr\"odinger type equation. In the case of a local quasipotential $V \left( \vec r \right)$ the equation for the wavefunction of spinless particles with equal masses has the form

\begin{equation}
\label{1}
\left[ {H_0  + V\left( {\vec r} \right)} \right]\psi \left( {\vec r} \right) = E\psi \left( {\vec r} \right).
\end{equation}

The free Hamiltonian operator $H_0$ is the finite-difference operator with the step equal to the Compton wavelength of the particle $\lambdabar=\hbar / mc$, i.e.

\begin{equation}
\label{2}
H_0  = mc^2 \left[ \cosh i \lambdabar \partial _r  + \frac{i \lambdabar}{r}\sinh i \lambdabar \partial _r  - \frac{\lambdabar^2 {\Delta _{\vartheta ,\varphi } }}{{2r^2 }}e^{i \lambdabar \partial _r } \right] ,
\end{equation}
where $\Delta _{\vartheta ,\varphi }$ is an angular part of the Laplace operator and $\partial _r=\partial / \partial r$. The quasipotential $V\left( {\vec r} \right)$ can be calculated in the framework of the quantum field theory, or introduced phenomenologically.

The space of three-dimensional vectors $\vec r$ is called the relativistic configurational space or $\vec r$-space. The transformation between configurational $\vec r$- and its canonically conjugate three-dimensional momentum $\vec p$-space is given by the kernel

\begin{eqnarray}
\label{3}
\xi \left( {\vec p,\vec r} \right) = \left( \frac {p_0  - \vec p\vec n} {mc} \right)^{ - 1 - ir/\lambdabar} ,
\\
\vec r=r\vec n, \; \vec n^2=1, \; 0\leq r < \infty, \nonumber
\end{eqnarray}
rather than by the Fourier kernel $\exp \left( {i\vec p\vec r /\hbar} \right)$ in the non-relativistic case.

The relativistic plane wave (\ref{3}) is the generating function for the matrix elements of the unitary irreducible representations of the Lorentz group $SO \left( 3,1 \right)$. The momenta of particles belong to the upper sheet of the mass hyperboloid $p_0 ^2  - \vec p^2  = m^2 c^2$, and form a three-dimensional Lobachevsky space, whose group of motion is the Lorentz group. The functions (\ref{3}) are eigenfunctions of the Hamiltonian $H_0$ (\ref{2}), i.e.

\begin{equation}
\label{4}
\left( {H_0  - E_p } \right)\xi \left( {\vec p,\vec r} \right) = 0,\;E_p  = cp_0  = c \sqrt {m^2 c^2 + \vec p^2 } 
\end{equation}
and in the non-relativistic limit they coincide with the Euclidean plane waves

\begin{equation}
\label{5}
\mathop {\lim }\limits_{c \to \infty } \xi \left( {\vec p,\vec r} \right) = e^{i\vec p\vec r /\hbar} .
\end{equation}

In the case of $O(3)$ symmetrical quasipotentials $V\left( {\vec r} \right) = V\left( r \right)$ (\ref{1}), the angular dependences of the wavefunction $\psi \left( {\vec r} \right)$ are separated in the standard manner

\begin{equation}
\label{6}
\psi \left( {\vec r} \right) = \frac{1}{r}R_l \left( r \right)Y_{lm} \left( {\vartheta ,\varphi } \right),
\end{equation}
where $l = 0,1,2, \ldots$ is the orbital quantum number. The radial wavefunction $R_l (r)$ satisfies the equation

\begin{equation}
\label{7}
\left[ {\tilde H_0 ^{rad}  + \tilde V\left( r \right)} \right]R_l \left( r \right) = E_l R_l \left( r \right),
\end{equation}
where

$$
\tilde H_0 ^{rad}  = mc^2 \left[ \cosh i \lambdabar \partial _r  + \frac{{l\left( {l + 1} \right)}}{{2\left(r/\lambdabar\right)^{\left( 2 \right)} }}e^{i \lambdabar \partial _r } \right],\;\tilde V\left( r \right) = rV\left( r \right)\frac{1}{r},
$$
and $\left(r/\lambdabar\right)^{\left( \delta  \right)}  = i^\delta  \frac{{\Gamma \left( { - ir / \lambdabar + \delta } \right)}}{{\Gamma \left( { - ir/ \lambdabar} \right)}}$ is a generalized degree \cite{kadyshevsky2}.

\section{The $N$-dimensional case}

The $N$-dimensional relativistic configurational $\vec r_N$-space is introduced in \cite{atakishiyev} by analogy with the three-dimensional relativistic configurational $\vec r$-space \cite{kadyshevsky}. Namely, transition to the $\vec r_N$-space

\begin{equation}
\label{8}
\psi \left( {\vec r_N } \right) = \frac{mc}{{\left( {2\pi \hbar} \right)^{N/2} }}\int {\frac{{d^N p}}{{p_0 }}\xi \left( {\vec p_N ,\vec r_N } \right)\psi \left( {\vec p_N } \right)} 
\end{equation}
is performed by the expansion in terms of the matrix elements

\begin{eqnarray}
\label{9}
\xi \left( {\vec p_N ,\vec r_N } \right) = \left( \frac {p_0  - \vec p_N \vec n_N } {mc} \right)^{ - \frac{{N - 1}}{2} - ir/ \lambdabar} ,
\\
\vec p_N  = \left( {p_1 ,p_2 , \ldots p_N } \right),\;\vec n_N  = \left( {n_1 ,n_2 , \ldots n_N } \right),\;\vec r_N  = r\vec n_N ,\;\vec n_N ^2  = 1,
\nonumber
\end{eqnarray}
of the infinite-dimensional unitary representations of the $N$-dimensional momentum Lobachevsky space motion group $SO\left( {N,1} \right)$, which is realized on the upper sheet of the hyperboloid $p_0 ^2  - \vec p_N ^2  = p_0 ^2  - p_1 ^2  - p_2 ^2  -  \cdots - p_N ^2  = m^2 c^2$.

The $N$-dimensional plane waves obey the completeness and orthogonality conditions

\begin{eqnarray}
\label{10}
 \frac{1}{{\left( {2\pi \hbar} \right)^{N/2} }}\int {\frac{{d^N p}}{{p_0 }}\xi ^* \left( {\vec p_N ,\vec r_N } \right)\xi \left( {\vec p_N ,\vec r'_N } \right)}  = w_N ^{ - 1} \left( r \right)\delta \left( {\vec r_N  - {\vec r_N}' } \right), \\ 
 \frac{mc}{{\left( {2\pi \hbar } \right)^{N/2} }}\int {w_N \left( r \right)d^N  r \xi ^* \left( {\vec p_N ,\vec r_N } \right)\xi \left( {\vec p'_N ,\vec r_N } \right)}  = p_0 \delta \left( {\vec p_N  - \vec p'_N } \right), 
\end{eqnarray}
where $w_N \left( r \right) = \left( r/ \lambdabar \right)^{1 - N}  \left| {\left( r/ \lambdabar \right)^{\left( {\frac{{N - 1}}{2}} \right)} } \right|^2$.

The functions (\ref{9}) are the eigenfunctions of the free Hamiltonian $H_0 ^{\left( N \right)}$, which in the spherical system of coordinates $\vec r_N  = \left( {r,\vartheta _1 ,\vartheta _2 , \ldots ,\vartheta _{N - 1} } \right)$ has the form

\begin{equation}
\label{11}
H_0 ^{(N)}  = mc^2 \left\{ \cosh i\lambdabar \partial _r  + \frac{{i\lambdabar (N - 1)}}{{2r}}\sinh i \lambdabar \partial _r  - \frac{{\lambdabar ^2 \Delta _0 ^{(N)} }}{{r\left[ {2r - i \lambdabar \left( {N - 3} \right)} \right]}}e^{i \lambdabar \partial _r } \right\}. \quad 
\end{equation}

Here $\Delta _0 ^{(N)}$ is an angular part of the $N$-dimensional Laplace operator. The eigenfunctions of $\Delta _0 ^{(N)}$ are the $N$-dimensional spherical harmonics \cite{erdelyi}:

\begin{equation}
\label{12}
\Delta _0 ^{\left( N \right)} Y_N ^l \left( {\vartheta _1 ,\vartheta _2 , \ldots ,\vartheta _{N - 1} } \right) =  - l\left( {l + N - 2} \right)Y_N ^l \left( {\vartheta _1 ,\vartheta _2 , \ldots ,\vartheta _{N - 1} } \right).
\end{equation}

In the non-relativistic limit Hamiltonian (\ref{11}) coincides with the non-relativistic free Hamiltonian, i.e.

\begin{equation}
\label{13}
\mathop {\lim }\limits_{c \to \infty } \left( {H_0 ^{\left( N \right)}  - mc^2} \right) =  - \frac{\hbar ^2}{2m}\Delta ^{\left( N \right)}  =  - \frac{\hbar^2}{2m}\left( {\partial _r ^2  + \frac{{N - 1}}{r}\partial _r  + \frac{{\Delta _0 ^{\left( N \right)} }}{{r^2 }}} \right).
\end{equation}

The interacting particles are described by the equation

\begin{equation}
\label{14}
\left[ {H_0 ^{\left( N \right)}  + V\left( {\vec r_N } \right)} \right]\psi \left( {\vec r_N } \right) = E_N \psi \left( {\vec r_N } \right).
\end{equation}

The free Hamiltonian $H_0 ^{\left( N \right)}$ (\ref{11}) is invariant with respect to the $N$-dimensional rotation group $O\left( N \right)$. Thus, for the central quasipotentials $V\left( {\vec r_N } \right) = V\left( r \right)$ the wavefunction depends on the spherical angles ${\vartheta _1 ,\vartheta _2 , \ldots ,\vartheta _{N - 1} }$ in the standard manner

\begin{equation}
\label{15}
\psi _{Nl} \left( {\vec r_N } \right) = \psi _{Nl} \left( r \right) \cdot Y_N ^l \left( {\vartheta _1 ,\vartheta _2 , \ldots ,\vartheta _{N - 1} } \right)
\end{equation}
and therefore an $N$-dimensional problem is reduced to finding the eigenvalues and eigenfunctions of the radial part of a Hamiltonian

\begin{eqnarray}
\label{16}
 \left[ {H_0 ^{\left( N \right)\;rad}  + V\left( r \right)} \right]\psi _{Nl} \left( r \right) = E_{Nl} \psi _{Nl} \left( r \right), \\ 
 H_0 ^{\left( N \right)\;rad}  = mc^2 \left\{ \cosh i \lambdabar \partial _r  + \frac{{i \lambdabar \left( {N - 1} \right)}}{{2r}}\sinh i \lambdabar \partial _r  + \frac{{\lambdabar ^2 l\left( {l + N - 2} \right)}}{{r\left[ {2r - i \lambdabar \left( {N - 3} \right)} \right]}}e^{i \lambdabar \partial _r } \right\} . \nonumber
\end{eqnarray}

Due to the appearance of the weight function ${w_N \left( r \right)}$, 
\begin{equation}
\label{17}
w_N \left( r \right)d^N \vec r_N  = \lambdabar^{N-1} \left| {\left(r/ \lambdabar \right)^{\left( {\frac{{N - 1}}{2}} \right)} } \right|^2 drd\Omega _{N - 1} ,
\end{equation}
in the ${\vec r_N }$-space orthogonality condition (\ref{10}), inclusion of the multiplier $\left[ {\left( { - r} / \lambdabar \right)^{\left( {\frac{{N - 1}}{2}} \right)} } \right]^{ - 1}$ to the expression of the wavefunction

\begin{equation}
\label{18}
\psi _{Nl} \left( r \right) = \left[ {\left( { - r} / \lambdabar \right)^{\left( {\frac{{N - 1}}{2}} \right)} } \right]^{ - 1} R_{Nl} \left( r \right)
\end{equation}
allows us to reduce $N$-dimensional radial equation (\ref{16}) to the simplest form

\begin{equation}
\label{19}
\left[ {\tilde H_0 ^{\left( N \right)\;rad}  + \tilde V\left( r \right)} \right]R_{Nl} \left( r \right) = E_{Nl} R_{Nl} \left( r \right),
\end{equation}
where

\begin{eqnarray}
\label{20}
 \tilde H_0 ^{\left( N \right)\;rad}  = \left( { - r} / \lambdabar \right)^{\left( {\frac{{N - 1}}{2}} \right)} H_0 ^{\left( N \right)\;rad} \left[ {\left( { - r} / \lambdabar \right)^{\left( {\frac{{N - 1}}{2}} \right)} } \right]^{ - 1}  = mc^2 \left[ \cosh i \lambdabar \partial _r  + \frac{{L\left( {L + 1} \right)}}{{2\left(r / \lambdabar \right)^{(2)} }}e^{i \lambdabar \partial _r } \right], \nonumber \\ 
 \tilde V\left( r \right) = \left( { - r} / \lambdabar \right)^{\left( {\frac{{N - 1}}{2}} \right)} V\left( r \right)\left[ {\left( { - r} / \lambdabar \right)^{\left( {\frac{{N - 1}}{2}} \right)} } \right]^{ - 1} , \quad \quad \quad \\ 
 L = l + \left( {N - 3} \right)/2. \quad \quad \quad \nonumber
\end{eqnarray}

Eq. (\ref{19}) coincides by form with the three-dimensional radial finite-difference equation (\ref{7}).

\section{The relativistic $N$-dimensional model}

Let us consider a model of the relativistic $N$-dimensional singular oscillator, which corresponds to the following choice of the quasipotential

\begin{equation}
\label{21}
V\left( r \right) = \left\{ {\frac{1}{2} m \omega ^2 \frac{{r\left( {r + i \lambdabar} \right)^2 }}{{r - i \lambdabar \left( {N - 3} \right)/2}} + \frac{g}{{r\left[ {r - i \lambdabar \left( {N - 3} \right)/2} \right]}}} \right\}e^{i \lambdabar \partial _r } .
\end{equation}

In the non-relativistic limit we have

$$
\mathop {\lim }\limits_{c \to \infty } V\left( r \right) \to \frac{1}{2} m \omega ^2 r^2  + \frac{g}{{r^2 }}.
$$

In case of dimensionless variable $\rho=\frac x \lambdabar$ and parameters $\omega _0  = \frac{{\hbar \omega }}{{mc^2 }}$, $g_0  = \frac{{mg}}{{\hbar ^2 }}$ equation (\ref{19}) for the relativistic $N$-dimensional singular oscillator takes the form 

\begin{equation}
\label{22}
\left\{ {\cosh i\partial _\rho  + \frac{{L\left( {L + 1} \right)}}{{2\rho^{(2)} }}e^{i\partial _\rho }  + \left[ {\frac{1}{2}\omega_0 ^2 \rho^{\left( 2 \right)}  + \frac{g_0}{{\rho^{\left( 2 \right)} }}} \right]e^{i\partial _\rho } } \right\}R_{Nl} \left( \rho \right) = E_{Nl} R_{Nl} \left( \rho \right). \quad
\end{equation}

This equation was studied in \cite{nagiyev2}, where it has been shown that the radial wavefunctions $R_{Nl} \left( \rho \right)$ are expressed through the continuous dual Hahn polynomials $S_n \left( {x^2 ;a,b,c} \right)$ \cite{koekoek},

\begin{equation}
\label{23}
R_{Nnl} \left( \rho \right) = C_{Nnl} \left( { - \rho} \right)^{\left( {\alpha _L } \right)} \omega_0 ^{i\rho} \Gamma \left( {\nu _L  + i\rho} \right)S_n \left( {\rho^2 ;\alpha _L ,\nu _L ,\frac{1}{2}} \right),
\end{equation}
where $n=0,1,2,3, \dots$ and

\begin{eqnarray}
\label{24}
\alpha _L  = \frac{1}{2} + \frac{1}{2}\sqrt {1 + \frac{2}{{\omega_0 ^2 }}\left( {1 - \sqrt {1 - 8g_0 \omega_0 ^2  - 4\omega_0 ^2 L\left( {L + 1} \right)} } \right)}, \\
\label{25}
\nu _L  = \frac{1}{2} + \frac{1}{2}\sqrt {1 + \frac{2}{{\omega_0 ^2 }}\left( {1 + \sqrt {1 - 8g_0 \omega_0 ^2  - 4\omega_0 ^2 L\left( {L + 1} \right)} } \right)}.
\end{eqnarray}

The eigenvalues of Hamiltonian (\ref{22}) corresponding to the wavefunctions are

\begin{equation}
\label{26}
E_{Nnl}  = \hbar \omega \left( {2n + \alpha _L  + \nu _L } \right), \; n=0,1,2,3, \dots.
\end{equation}

The normalization coefficients $C_{Nnl}$ in (\ref{23}) are defined by the condition

\begin{equation}
\label{27}
\int\limits_0^\infty  {R_{Nnl}^* \left( \rho \right)R_{Nml} \left( \rho \right)dr}  = \delta _{nm} .
\end{equation}

\section{A dynamical symmetry}

It turns out that the Hamiltonian

\begin{equation}
\label{28}
\tilde H^{\left( N \right)rad}  = \hbar \omega \left( {a^ +  a^ -   + \alpha _L  + \nu _L } \right)
\end{equation}
may be factorized in terms of the difference operators \cite{nagiyev1}

\begin{eqnarray}
\label{29}
 a^ -   = \frac{1}{{\sqrt {2\omega_0 } }}\left[ {e^{ - \frac{i}{2}\partial _\rho }  - \omega_0 e^{\frac{i}{2}\partial _\rho } \left( {\nu _L  + i\rho} \right)\left( {1 + \frac{{\alpha _L }}{{i\rho}}} \right)} \right], \nonumber \\ \\ 
 a^ +   = \frac{1}{{\sqrt {2\omega_0 } }}\left[ {e^{ - \frac{i}{2}\partial _\rho }  - \omega_0 \left( {\nu _L  - i\rho} \right)\left( {1 - \frac{{\alpha _L }}{{i\rho}}} \right)e^{\frac{i}{2}\partial _\rho } } \right]. \nonumber 
 \end{eqnarray}

They are a pair of Hermitian conjugate operators. Using $a ^-$ and $a ^+$, one can construct a dynamical symmetry algebra of the relativistic $N$-dimensional singular oscillator under consideration. The result is as follows \cite{nagiyev1}.

The lowering and raising operators are

\begin{eqnarray}
\label{30}
A^ -   = \frac{1}{{2\omega_0 }}\left[ {\left( {\omega_0 \rho + \frac {iP}{mc}} \right)^2  - \frac{{2g_0+L \left(L+1 \right)}}{{1 + \rho^2 }}} \right],\nonumber \\ \\ A^ +   = \frac{1}{{2\omega_0 }}\left[ {\left( {\omega_0 \rho - \frac {iP}{mc}} \right)^2  - \frac{{2g_0+L \left(L+1 \right)}}{{1 + \rho^2 }}} \right], \nonumber
\end{eqnarray}
where a generalized momentum operator is

\begin{equation}
\label{31}
P = \frac i c \left[ {\tilde H^{\left( N \right)rad} ,\rho} \right] =  - mc \left[ \sinh i\partial _\rho  - \left( {\frac{1}{2}\omega_0 ^2 \rho^{\left( 2 \right)}  + \frac{g_0 + L \left(L+1 \right)/2}{{\rho^{\left( 2 \right)} }}} \right)e^{i\partial _\rho } \right] .
\end{equation}

They satisfy the following commutation relations:

\begin{eqnarray}
\label{32}
 \left[ {\tilde H^{\left( N \right)rad} ,A^ \pm  } \right] =  \pm 2 \hbar \omega A^ \pm  , \\ 
\label{33}
 \left[ {A^ -  ,A^ +  } \right] = \frac {\omega_0}{mc^2} \tilde H^{\left( N \right)rad} \left\{ {1 + \frac{2}{{\omega_0 ^2 }}\left[ {\left( \frac {\tilde H^{\left( N \right)rad} } {mc^2} \right)^2  - 1} \right]} \right\}. 
\end{eqnarray}

Now we define a set of operators

\begin{equation}
\label{34}
K^ -   = A^ -  f^{ - 1/2} \left( {\tilde H^{\left( N \right)rad} } \right),\;K^ -   = f^{ - 1/2} \left( {\tilde H^{\left( N \right)rad} } \right)A^ -  ,\;K_0  = \frac{1}{{2\hbar \omega }}\tilde H^{\left( N \right)rad} ,
\end{equation}
where $f\left( x \right) = \left[ {x/mc^2 + \omega_0 \left( {\alpha _L  - \nu _L  - 1} \right)} \right]\left[ {x/ mc^2 + \omega_0 \left( {\nu _L  - \alpha _L  - 1} \right)} \right]$.

These operators provide a realization of the $SU\left( {1,1} \right)$ Lie algebra on the basis elements $\left\{ {R_{Nnl} } \right\}_{\scriptstyle n = 0 \hfill \atop 
  \scriptstyle l = 0 \hfill}^\infty  $, i.e.

\begin{equation}
\label{35}
\left[ {K_0 ,K^ \pm  } \right] =  \pm K^ \pm  ,\;\left[ {K^ -  ,K^ +  } \right] = 2K_0 .
\end{equation}

The operator $K_0$, which generates a compact subgroup of $SU\left( {1,1} \right)$ has a discrete spectrum in a $D^ +  \left( s \right)$ representation, which is bounded below and is equal to $n+s$, where $n = 0,1,2, \ldots$, and $s$ is the so-called Bargmann index, $s>0$.

With $K^2  = K_0 \left( {K_0  - 1} \right) - K^ +  K^ -  $ as the Casimir operator, one has $K^2  = s\left( {s - 1} \right)$. For the operators (\ref{34}) one has $K^2  = \frac{1}{4}\left( {\alpha _L  + \nu _L } \right)\left( {\alpha _L  + \nu _L  - 2} \right)$, so that $s = \left( {\alpha _L  + \nu _L } \right)/2$. Thus we obtain algebrically the correct spectrum of the operator $\tilde H^{\left( N \right)rad}  = 2\hbar \omega K_0 $.

We note that the action of the operators $K^\pm$ on the wavefunctions $R_{Nnl} \left( r \right)$ is defined by the formulae

\begin{equation}
\label{36}
K^ +  R_{Nnl} \left( r \right) = \kappa _{n + 1} R_{N\left( {n + 1} \right)l} \left( r \right),\;K^ -  R_{Nnl} \left( r \right) = \kappa _n R_{N\left( {n - 1} \right)l} \left( r \right),
\end{equation}
where $\kappa _n  = \sqrt {n\left( {n + \alpha _L  + \nu _L  - 1} \right)}$.

From (\ref{36}) it follows that

\begin{equation}
\label{37}
R_{Nnl} \left( r \right) = \left[ {n!\left( {\alpha _L  + \nu _L } \right)_n } \right]^{ - 1/2} \left( {K^ +  } \right)^n R_{N0l} \left( r \right).
\end{equation}

\section{Conclusion}

There are a lot of interesting publications devoted to the study of the various charactiersics and properties of the non-relativistic $N$-dimensional isotropic harmonic oscillator \cite{george,calogero,bakhrakh,chang,kalnins,cardoso}. Various interesting approaches to generalize the non-relativistic oscillator exact-solvable models for the relativistic $N$-dimensional case also exist \cite{atakishiyev,ruijsenaars1,ruijsenaars2,vandiejen,braden,howard,odake}.

In this paper we generalized the relativistic finite-difference model of the istropic three-dimensional singular oscillator \cite{nagiyev2} to the $N$-dimensional case. It is exactly-solvable. We determined energy spectrum and wavefunctions of the problem and constructed a dynamical symmetry algebra. The finite-difference operators $K^+$, $K^-$ and $K_0$ generate $SU\left( {1,1} \right)$ group and $K_0$ has a discrete spectrum in a $D^ +  \left( s \right)$ representation, which is bounded below and is equal to $n+s$. 

It is necessary to note the recent published work \cite{odake2}, where the dynamical symmetry and factorization scheme is proposed for Hamiltonian leading to eigenfunctions expressed by continuous dual Hahn polynomials. The finite-difference Hamiltonian is formulated in the framework of Ruijsenaars-Schneider approach \cite{ruijsenaars1,ruijsenaars2,vandiejen} and therefore, it is different than factorization scheme used here and in \cite{nagiyev3}. Taking into account that, both of these approaches to solve explicitly the finite-difference equations lead to the close results, we aim to show in detail possible relations and transformations between them as a further step of investigations.

\section*{Acknowledgement}

One of the authors (E.I.J.) would like to acknowledge that this work is performed in the framework of the Fellowship 05-113-5406 under the INTAS-Azerbaijan YS Collaborative Call 2005.



\end{document}